\documentclass[preprint]{aastex}

\newcommand \mum{$\mu$m}
\newcommand \lsol{L$_{\odot}$}
\newcommand \msol{M$_{\odot}$}

\newfont{\rten}{cmr10}

\begin{document}

\slugcomment{To appear in {\it Astron.~J}; preprint--July 10, 2002}

\title{A Study of the Dynamics of Dust from the Kuiper Belt:
Spatial Distribution and Spectral Energy Distribution}

\author{Amaya Moro-Mart\'{\i}n\altaffilmark{1} and  Renu Malhotra\altaffilmark{2}}

\email{amaya@as.arizona.edu} 
\email{renu@lpl.arizona.edu}

\altaffiltext{1}{Steward Observatory, University of Arizona,
933 N. Cherry Av, Tucson, AZ 85721, USA}

\altaffiltext{2}{Department of Planetary Sciences, University of Arizona,
1629 E. University Boulevard, Tucson, AZ 85721, USA}

\begin{abstract}
The dust produced in the Kuiper Belt (KB) spreads throughout the Solar System 
forming a dust disk. We numerically model the orbital evolution of KB dust and 
estimate its equilibrium spatial distribution and its brightness and spectral
energy distributions (SED), assuming greybody absorption and emission by the dust grains. 
We show that the planets modify the KB disk SED, 
so potentially we can infer the presence of planets in spatially unresolved debris disks 
by studying the shape of their SEDs.
We point out that there are inherent uncertainties in the prediction of 
structure in the dust disk, owing to the chaotic dynamics of dust orbital 
evolution imposed by resonant gravitational perturbations of the planets.

\end{abstract}

\keywords{celestial mechanics --- interplanetary medium --- Kuiper Belt
--- methods: n-body simulations --- methods: numerical --- planetary systems
--- solar system: general} 

\section{Introduction}
Main sequence stars are commonly surrounded by cold far-IR-emitting material.
The fact that this infrared excess is not restricted to young stars, and that 
the dust grain removal processes, Poynting-Robertson (P-R) and solar wind 
drag, act on timescales
much smaller than the age of the system, indicate
that: (1) a reservoir of undetected dust-producing planetesimals exists; and 
(2) to induce frequent mutual collisions, their orbits must be dynamically 
perturbed by massive planetary bodies.
The Solar System is also filled with interplanetary dust. In the inner Solar
System, this dust, which gives rise to the zodiacal light, has been observed 
by Pioneer 10 (out to 3.3 AU) and by the infrared telescopes IRAS and 
COBE. The dominant sources of the zodiacal cloud are debris 
from Jupiter family short period comets and asteroids 
(Liou et al.,~\citeyear{liou95}; Dermott et al.,~\citeyear{derm92}).  
The discovery of a debris disk around $\beta$-Pictoris, 
extending to 100s of AU, together with the confirmation of the 
existence of the theoretically predicted Kuiper Belt objects 
(KBOs) (Jewitt \& Luu,~\citeyear{jewi95}), suggest that significant 
dust production may also occur in the outer Solar System due to mutual 
collisions of KBOs 
(Backman \& Paresce,~\citeyear{back93}; 
Backman, Dasgupta \& Stencel,~\citeyear{back95}; Stern,~\citeyear{ster96}) and 
collisions with interstellar grains (Yamamoto \& Mukai,~\citeyear{yama98}).

Dust particles are small enough to experience the effect of 
radiation and stellar wind forces. Radiation
pressure makes their orbital elements and specific orbital energy change 
immediately upon release from parent bodies.  If their orbital energy becomes
positive, the dust particles escape on hyperbolic orbits. 
In the Solar System, these particles are known as $\beta$-meteoroids 
(Zook \& Berg,~\citeyear{zook75}). If their orbital
energy remains negative, the dust particles stay on bound orbits.
P-R and solar wind drag tends to circularize and 
decrease the semimajor axis of these orbits, forcing these particles to slowly
drift in towards the central star (Burns, Lamy \& Soter,~\citeyear{burn79}). 
Assuming that the dust particles are constantly being produced, 
this drifting in creates a dust disk of wide radial extent, that we refer to as a 
$\it{debris~disk}$. Debris disks are systems that satisfy the following
conditions:
(1) their age is longer than the P-R and collisional lifetimes;
(2) they are optically thin to stellar radiation, even along the mid
plane; and 
(3) they have little or no gas, so that the dust dynamics is controlled by
gravitation and radiation forces only (Backman, \citeyear{back02}).

When planets are present, the journey of the dust particle towards the central star 
is temporarily interrupted by the trapping of the particle in Mean Motion 
Resonances (MMRs). MMRs occur when the orbital period of the 
particle is in a ratio of small integers to that of the perturbing planet.
[The p:q MMR means that the orbital period of the particle is p/q times 
that of the planet.] 
In an MMR, the drifting in is halted because the energy loss due to P-R drag 
is balanced by the resonant interaction with the planet's gravity field.
This trapping can potentially create structure in debris disks, as the particles
accumulate at certain semimajor axes.  
Sufficiently massive planets may also scatter and eject 
dust particles out of a planetary system, creating dust free or depleted 
zones. This structure, if observed, can be used to infer the presence of planets.
Liou \& Zook (\citeyear{liou99a}, hereafter LZ99) found that the presence of the 
Giant Planets has an important effect on the structure of the debris disk that
is presumably generated in the KB: 
Neptune creates a ring-like structure between 35 and 50 AU, due to the trapping of 
particles in exterior MMRs, and Jupiter and Saturn are responsible for the 
ejection of about 80\% of particles from the Solar System 
(Liou, Zook \& Dermott,~\citeyear{liou96}, hereafter LZD96). The latter 
creates a clearing in the inner 10 AU that resembles the inner gap in 
the $\beta$-Pictoris disk. If observed from afar,
the KB disk would be the brightest extended feature in the Solar System, 
and its structure,
if spatially resolved, could be recognized as harboring at least two giant planets: 
an inner planet (Jupiter plus Saturn) and outer planet (Neptune) (LZD96).
In anticipation of future observations of debris disks, whose structure is likely to be 
spatially unresolved, in this paper we are interested in
studying how the structure affects the shape of the disk SED and consequently 
if the SED can be used to infer the presence of planets.

In this paper we are going to follow numerically, from source to sink, the evolution of 
several hundred dust particles from the KB in the size range from 
1 to 40 $\mu$m (for $\rho$=2.7 g/cm$^{3}$), or from 3 to 120 $\mu$m (for $\rho$=1 g/cm$^{3}$),
under the combined effects of solar gravity, solar radiation pressure, P-R and solar wind
drag and the gravitational forces of 7 planets (excluding Mercury and Pluto). 
The sinks of dust included in our numerical simulations are: (1) 
ejection into unbound orbits; (2) accretion onto the planets; and 
(3) orbital decay to less than 0.5 AU heliocentric distance.
The equations of motion are integrated using a modification of the multiple
time step symplectic method SyMBA (DLL98).
In $\S$2 we describe our numerical integration method and the tests performed to check the 
suitability of the code. $\S$3 describes our methods for deriving 
the equilibrium spatial distribution of the dust disk. 
$\S$4 explains the distribution of parent
bodies and the orbital evolution of dust. In $\S$5 we discuss
the formation of structure in the KB debris disk and its observational signatures.  
Dust destruction processes are discussed in $\S$6, and $\S$7 
summarizes our results.

\label{intro}

\section{The Numerical Method}
\label{num}
In order to study the dynamics of dust from the KB
we need to solve the problem of the dynamical evolution of micron-sized 
particles, under the effect of gravitational forces of the Sun and
the planets and radiation and solar wind forces. This has been solved in
the past using the adaptive step size Runge-Kutta integrator RADAU 
(LZD96; Liou \& Zook,~\citeyear{liou97}; Kortenkamp \& Dermott,~\citeyear{kort98};
LZ99 and Liou, Zook \& Jackson,~\citeyear{liou99b}). 
Another possible choice is the 
standard mixed variable symplectic (MVS) integrator, developed by Wisdom \& Holman 
(\citeyear{wisd91}). Its advantage over implicit Runge-Kutta 
integrators is its speed, about an order of magnitude faster (Wisdom \& Holman,
~\citeyear{wisd91}).
This is why the MVS method is now
used in long-term studies of the Solar System, allowing to reach integration 
times approaching the age of the system. Its disadvantage, however, 
is that it cannot handle close encounters amongst bodies. Since 
the outcome of close encounters between the dust particle and the planets
is critical for the study of the dynamical evolution of dust grains, previous
researchers have chosen RADAU as their numerical integrator.
But recently, Duncan, Levison \& Lee (\citeyear{dunc98}; hereafter DLL98) 
have developed a new multiple time step symplectic algorithm, SyMBA, that can 
handle close encounters in a sympletic way, thus retaining the speed of the 
MVS method while being able to overcome its main disadvantage.

The equations of motion of the N-body system are integrated
using a variation of SyMBA called SKEEL, 
which we have modified to include ratiation forces.
In this section, we summarize the main features of SKEEL as described in DLL98, 
followed by a description of how radiation forces were introduced and the 
tests that we have performed to check the validity of our results.

\subsection{The Multiple Time Step Symplectic Integrator SKEEL}
SKEEL solves the Newtonian gravitational N-body
problem by separating its Hamiltonian, 
\begin{equation}
{H(\bf{Q}_{\it{i}},\bf{P}_{\it{i}}) 
= \sum_{\it{i}=1}^{\it{n}} \left({\mid\bf{P}_{\it{i}}\mid^2 \over 2\it{m_{\it{i}}}} - {G\it{m_{i}} \it{m}_0 \over 
\mid\bf{Q}_{\it{i}}\mid} \right)+  {\mid\bf{P}_0\mid^2 \over 
2\it{m}_{tot}} + {1 \over 2\it{m}_0} \mid\sum_{\it{i}=1}^{\it{n}}\bf{P}_{\it{i}}\mid^2 - \sum_{\it{i}=1}^{\it{n}-1} \sum_{\it{j}=\it{i}+1}^{\it{n}} 
{G\it{m_{\it{i}}}\it{m_{\it{j}}} \over \mid \bf{Q}_{\it{i}}- \bf{Q}_{\it{j}}
\mid}}
\end{equation}
into three integrable components,
\begin{equation}
{H(\bf{Q}_{\it{i}},\bf{P}_{\it{i}})=H_{Kep}+ H_{Sun}+ H_{int}},
\end{equation}
where
\begin{equation}
{H_{Kep}=\sum_{\it{i}=1}^{\it{n}} \left({\mid\bf{P}_{\it{i}}\mid^2 \over 2\it{m_{\it{i}}}} - {G\it{m_{i}} \it{m}_0 \over 
\mid\bf{Q}_{\it{i}}\mid} \right)},
\end{equation}
\begin{equation}
{H_{Sun}={1 \over 2\it{m}_0} \mid\sum_{\it{i}=1}^{\it{n}}\bf{P}_{\it{i}}\mid^2},
\end{equation}
\begin{equation}
{H_{int}= - \sum_{\it{i}=1}^{\it{n}-1} \sum_{\it{j}=\it{i}+1}^{\it{n}} 
{G\it{m_{\it{i}}}\it{m_{\it{j}}} \over \mid \bf{Q}_{\it{i}}- \bf{Q}_{\it{j}}
\mid}.}
\end{equation}
$\bf{Q}_{\it{i}}$ and $\bf{P}_{\it{i}}$ are respectively the heliocentric positions and barycentric 
momenta (if $\it{i} \neq$ 0) and the position of the center of mass and the 
total momentum of the system (if $\it{i}$~= 0). $\it{m}_{tot}$ is the total
mass of the system. The contribution from the second term in the $\it{rhs}$ of (1) is 
ignored because it corresponds to the free motion of the center of mass.
A second order symplectic integrator consists in approximating the time evolution by
the following symmetrized sequence of steps
\begin{equation}
{E_{Sun}\left({\tau \over 2} \right)E_{int}\left({\tau \over 2} \right)E_{Kep}(\tau) E_{int}\left({\tau \over 2} \right)E_{Sun}\left({\tau \over 2} \right)},
\end{equation}
where E$_{i}$($\tau$) is the evolution under H$_{i}$ for time $\tau$.
For each body there is: (1) a linear drift in position
by ($\tau$/2$\it{m_0}$) $\sum \bf{P}_{\it{i}}$, to account for the 
motion of the Sun with respect to the barycenter; (2) a kick to its 
momentum for time ($\tau$/2), to account for the gravitational
forces of all the massive bodies except the Sun; (3) an evolution
along a Kepler orbit for time $\tau$; (4) another kick like (2); (5) another
linear drift like (1).
During a close encounters between a particle and a planet, 
the contribution from the encountering planet is 
separated from the rest so that the time evolution becomes
\begin{equation}
{E_{Sun}\left({\tau \over 2} \right)E_{int}^{ne}\left({\tau \over 2} \right)E_{int}^{enc}\left({\tau \over 2}\right)
E_{Kep}(\tau) E_{int}^{enc}\left({\tau \over 2}\right) E_{int}^{ne}\left({\tau \over 2} \right)E_{Sun}\left({\tau \over 2} \right)},
\end{equation}
where E$_{int}^{ne}$ refers to the contribution to
H$_{int}$ from all the planets except the one in the encounter, and 
E$_{int}^{enc}$ is the same but for the planet in the encounter only.
The close-encounter algorithm, represented by
\begin{equation}
{E_{int}^{enc}\left({\tau \over 2}\right)
E_{Kep}(\tau) E_{int}^{enc}\left({\tau \over 2}\right),}
\end{equation}
is as follows. The two-body potential terms in H$_{int}$, due to the 
encountering planet, are decompossed into
\begin{equation}
{{G\it{m_{\it{i}}}\it{m_{\it{j}}} \over \mid \bf{Q}_{\it{i}}- \bf{Q}_{\it{j}}
\mid} = \sum_{\it{k}=0}^{\infty} V_{\it{k}}.}
\end{equation}
For details about the conditions V$_{k}$ need to satisfy and the particular 
functions used in SKEEL, see DLL98. The multiple time step method consists then in
applying (8) recursively, 
\begin{equation}
{E_{\Sigma_{0}}(\tau) \approx 
E_{0}\left({\tau_{0} \over 2}\right)
E_{\Sigma_{1}}(\tau_{0})
E_{0}\left({\tau_{0} \over 2} \right)
\approx E_{0}\left({\tau_{0} \over 2}\right)
[E_{1}\left({\tau_{1} \over 2}\right)
E_{\Sigma_{2}}(\tau_{1})
E_{1}\left({\tau_{1} \over 2}\right)]^M
E_{0}\left({\tau_{0} \over 2}\right)
}
\end{equation}
where E$_{\it_{i}}(\tau)$ and E$_{\Sigma_{i}}(\tau)$ 
are the evolution for time $\tau$ under V$_{i}$ and 
H$_{Kep}+H_{Sun}$ + $\Sigma_{\it{k}=i}^{\infty}$V$_{\it{k}}$ respectively. 
At each level of recursion, the evolution under
E$_{\Sigma_{i}}(\tau)$, is approximated by: (1) evolution under 
V$_{i}$ for $\tau$/2; (2) M second-order steps of length $\tau$; 
(3) evolution under V$_{i}$ for $\tau$/2.
This is equivalent to placing concentric shells around the massive
body; the smaller the shell, the smaller the time step associated with it,
allowing to resolve peri-planet passage.
In particular, DLL98 uses $\tau_{\it{k}}$/$\tau_{\it{k}+1}$ = M; for our 
runs, M = 3. 
Note that this multiple time step algorithm only activates during 
close encounters. When the bodies are farther apart, the algorithm
reduces to (6); this is because \{H$_{Sun}$,H$_{Kep}$\}=\{H$_{Sun}$,H$_{int}$\}=0, so that the pairs are interchangeable. 

We use units in which $\it{G}$=1; the unit of mass is 1\msol,
the unit of length is 1 AU and the unit of time is the period of a massless
particle at 1 AU divided by 2$\pi$.

\subsection{Radiation Pressure, Poynting-Robertson and Solar Wind Drag}
A particle of mass $\mu$ and geometric cross section $\it{A}$, at heliocentric position $\bf{r}$, 
moving with velocity $\bf{v}$ with respect to a central body of mass $\it{m}_0$, 
which is the source of a radiation field of  energy flux density $\it{S}$=L/4$\pi \it{r}^{2}$, 
$\it{feels}$ a force due to the absorption and re-emission of radiation that is given 
(to terms of order $\it{v}$/c) by 
\begin{equation}
{{d^2{\bf{r}} \over dt^2} = {-G\it{m}_0(1-\beta) \over \it{r}^{3}} {\it{\bf{r}}} 
- {\beta_{sw} \over c} {G \it{m}_0 \over \it{r}^{2} }\left[\left({\dot{r}\over r}\right)
{\bf{r}} + {\bf{v}}\right],
}
\end{equation}
where $\beta$ is a dimensionless constant equal to the ratio between 
the radiation pressure force, $\it{F_{r}}$=$\it{SAQ}_{pr}$/c, and 
the gravitational force, $\it{F}_{g}$=$\it{G\it{m}_0\mu}$/$\it{r}^{2}$, 
so that for spherical grains $\beta$ =$\it{F}_{r}$/$\it{F}_{g}$=$\it{SAQ}_{pr}\it{r}^{2}$/
($\it{G\it{m}_0\mu c}$)=(3L/16$\pi\it{G\it{m}_0c})(Q_{pr}$/$\rho$~s). For the Sun, 
$\beta$=5.7 $\times$ 10$^{-5}$ Q$_{pr}$/$\rho$ s, where $\rho$ and s are the
density and radius of the grain in cgs units (Burns, Lamy \& Soter,
~\citeyear{burn79}). Q$_{pr}$ is the 
radiation pressure coefficient, a measure of the fractional
amount of energy scattered and/or absorbed by the grain. Q$_{pr}$ is a 
function of the physical properties of the grain and the wavelength of
the incoming radiation; the value we use is an average integrated 
over the solar spectrum. The advantage of using the dimensionless
parameter $\beta$ is that it is independent of distance, being a function
only of the particle size and composition. $\beta_{sw}$ = (1+sw)$\beta$, where
$\it{sw}$ is the ratio of the solar wind drag to the P-R drag; in this paper 
we use a constant value $\it{sw}$=0.35 (Gustafson,~\citeyear{gust94}).

The Hamiltonian associated with the first term in the $\it{rhs}$ of 
(11) is H$_{Kep}$ in eq. (3), with $\it{m}_0$(1-$\beta$) 
instead of $\it{m}_0$.  Physically, this means that radiation pressure
makes the dust grain $\it{feel}$  a less massive Sun. In our numerical 
integrator, SKEEL-RAD, we introduce the second term
in (11), the P-R and solar wind drag term, as an additional 
$\it{kick}$ to the momentum of the particle. The algorithm thus becomes,
\begin{equation}
{E_{Sun}\left({\tau \over 2} \right)E_{int}^{rad}\left({\tau \over 2} \right)
E_{Kep}^{rad}(\tau)E_{int}^{rad}\left({\tau \over 2} \right)E_{Sun}\left({\tau \over 2} \right)}.
\end{equation}
In the inertial reference frame, the P-R drag can be thought of as a 
$\it{mass~loading~drag}$: the re-emitted radiation emits more momentum 
into the forward direction of motion due to the Doppler 
effect, which means that the particle loses momentum; since 
the mass is conserved, the particle is decelerated (there is a drag force).
In the particle's reference frame it originates from the aberration of 
the radiation, that generates a drag force.

\subsection{Comparison with Analytical Results}
There is no analytic solution to the general problem of a 
particle moving under the effect of gravitational forces from the Sun and the 
planets and radiation and solar wind forces.  For this reason, the code
cannot be tested in the most general case. But there are analytic solutions
for the evolution of the orbital elements of a particle under
the effect of radiation in the 2-body problem 
(Wyatt \& Whipple,~\citeyear{wyat50}; Burns, Lamy \& Soter,
~\citeyear{burn79}) and in the circular restricted 
3-body problem (Liou and Zook,~\citeyear{liou97}). We will use these 
solutions to test the numerical procedure and the validity of our results.

\subsubsection{Jacobi Constant Conservation}
In the circular restricted 3-body problem, consisting of a massless particle,
a central mass and a planet in a circular orbit, the Jacobi constant is
an integral of the motion. We have integrated the orbits of 
50 massless particles in the presence of the Sun and Neptune 
(with $\it{a}$=30 AU and $\it{e}$=0). The semimajor axes of the 
particles were uniformly distributed between 36 and 
40 AU and the perihelion distance was set to 30 AU. We use a step size of 
2 years and an integration time of 10$^9$ years. We found that 
34 out of 50 particles have close encounters, 
with $\Delta$J/J(0) $\sim$ O(10$^{-6}$)--O(10$^{-7}$). The remaining 
16 that do not suffer close encounters have 
$\Delta$J/J(0) $\sim$ O(10$^{-8}$).
The worst jacobi conservation has $\Delta$J/J(0) $\sim$ 7$\cdot$10$^{-6}$.
These results suggests that close encounters are integrated accurately.

\subsubsection{Rates of Change of Orbital Elements}
Burns, Lamy \& Soter (\citeyear{burn79}), following
Wyatt \& Whipple (\citeyear{wyat50}), derived the time rates of change
(averaged over an orbit) of semimajor axis ($\it{a}$), eccentricity ($\it{e}$), 
and inclination ($\it{i}$), of a 
particle in the 2-body problem in the presence of radiation and solar wind 
forces,
\begin{equation}
{({ da \over dt})_{PR}=-{(1+sw)\beta m_0 \over c}{2+3e^2 \over a(1-e^2)^{3/2}},}
\end{equation}
\begin{equation}
{({ de \over dt})_{PR}=-{5(1+sw)\beta m_0 \over 2c}{e \over a^2(1-e^2)^{1/2}},}
\end{equation}
\begin{equation}
{({ di \over dt})_{PR}=0}.
\end{equation}
Figure 1 presents the evolution of $\it{a}$ and $\it{e}$ for a particle
with $\beta$=0.2 and sw=0.35. The agreement between the numerical and 
analytical results is perfect.

When radiation is introduced into the circular restricted 3-body
problem, the Jacobi constant is no longer an integral of the motion.
Using the time variation of the Jacobi constant due to radiation and
solar wind forces, together with the time rate 
of change of the Tisserand criterion, Liou and Zook (\citeyear{liou97}) 
have derived analytic expressions that describe the orbital evolution of a 
particle trapped in a MMR with a planet. The equation 
relating the time variation in $\it{e}$ and $\it{i}$ is
\begin{equation}
{e(1-e^2)^{-1/2}\textrm{cos}i{de \over dt} +(1-e^2)^{1/2}\textrm{sin}i
{di \over dt}={(1+sw)\beta m_0 \over a^2 c}[\textrm{cos}i - 
{{a_{pl}}^{3/2}(3e^2+2)(1-\beta)^{1/2} \over 2a^{3/2}(1-e^2)^{3/2}}],}
\end{equation}
where $\it{a}$ and $\it{a}_{pl}$ are the semimajor axis of the 
resonant orbit and the planet respectively, related by equation
\begin{equation}
{a=a_{pl}(1-\beta)^{1/3}({p \over q})^{2/3}}.
\end{equation}
In the particular case when i=0,
\begin{equation}
{\dot e ={(1+sw)\beta m_0(1-e^2)^{1/2} \over a^2ce}[1-{
{a_{pl}}^{3/2}(3e^2+2)(1-\beta)^{1/2} \over 2a^{3/2}(1-e^2)^{3/2}}].}
\end{equation}
The expansion of eq. (16) to second order in $\it{e}$ and $\it{i}$ allows
to decouple  their time variations; after integrating the resulting 
two differential
equations, Liou and Zook (\citeyear{liou97}) arrive at these equations 
(valid only for e-type resonances),
\begin{equation}
{e^2=[{e_0}^2-{K-1 \over 3}] \textrm{exp}(-{3A \over K}t) + {K-1 \over 3},}
\end{equation}
\begin{equation}
{i=i_0 \textrm{exp}(-{A \over 4}t),}
\end{equation}
where A=2(1+sw)$\beta$ m$_0$/a$^2$c and K=p/q; p and q are the 
two integers that specify the p:q resonance (K$>$1 for exterior MMR, and
K$<$1 for interior MMR). To carry out the comparison between 
analytical and numerical results, we have followed the orbital evolution of 
100 pyroxene dust particles, 1\mum~ in diameter ($\beta$=0.17, sw=0.35) in
a circular Sun-Neptune system. The different panels in Figure 2 
show the evolution of four of these particles trapped in the 
1:1, 5:3, 4:3 and 5:6 MMR with Neptune ($\it{a}$=30 AU, $\it{e}$=0). 
The agreement with equation (19) is good 
for small eccentricities, where the analytical expression holds. 
We conclude that the code is treating radiation and solar wind forces accurately.

\section{Equilibrium Distribution}
\label{accum}
Ideally, one would like to be able to follow the evolution 
of a number of particles large enough to resolve the disk structure.
However, even though our numerical integrator is very 
efficient, this task is not feasible with the current computational power.
We estimate that for a 50 AU radius disk about 10$^5$ particles would be needed to 
resolve the structure induced by the Solar System planets. To get around this problem, 
LZ99 used the following approach to obtain the equilibrium spatial distribution 
of the dust using only 100 particle simulations: first integrate the orbits 
from their source in the KB until they are either ejected from the 
Solar System or drifted into the Sun, recording the positions of the particles 
every 1000 years; then transform the particles' coordinates into a reference frame
rotating with the planet dominating the structure (Neptune); and finally 
accumulate all the rotated coordinates. This yields a time-weighted spatial distribution
of the 100 particles over their dynamical lifetime. It is equivalent to the 
actual spatial density distribution of KB dust provided: 
(1) the dust production rate is in equilibrium with the loss rate, and
(2) the dust particle dynamics is ergodic (i.e. the time-weighting reflects
the spatial density). 
LZ99 point out one limitation of this approach: it assumes the same
planetary configuration at the time of release of the dust particles.
There are, however, other more important limitations that were overlooked by LZ99. 
We consider these in detail because this is presently the only
feasible approach to the problem of structure formation in debris disks.

\subsection{Distribution of Particle Lifetimes}
Owing to the ergodic assumption, the debris disk structure obtained using
LZ99 approach is determined to a large extent by the longest lived particles, 
which represent only a very small fraction of the dust population.
The question is: are these particles anomalous, or are they part 
of a continuous distribution of lifetimes? In the case of 
anomalies, the structure would be dominated by the dynamics 
of a very small number of particles of uncertain significance, in which case the 
structure obtained by LZ99 approach would not necessarily resemble any equilibrium 
distribution. If the second option were true, however, the longest lived 
particles would indeed be statistically significant, since they will 
represent the contribution from an existing population of particles 
whose lifetimes are part of a long tail in a continuous distribution.

In order to answer this question, we have studied the lifetimes 
of the particles in all our models for the Solar System run so far. 
To facilitate discussion, we include here the list of our models.\\
- Models I-A: parent bodies at $\it{a}$=45 AU, 
$\it{e}$=0.1 and $\it{i}$=10$^\circ$. 4 giant planets. \\
- Models I-B: same as above but without planets.\\
- Models II-A: parent bodies randomly distributed between 
$\it{a}$=35-50AU, $\it{q}$=35-50 AU and $\it{i}$=0-17$^\circ$. 7 planets 
(excluding Mercury and Pluto).\\
- Models II-B: same as above but without planets. 

For all models, the mean anomaly 
(M), longitude of ascending node ($\Omega$) and  argument of perihelion 
($\omega$) of dust particles, were randomly distributed between 0 and 2$\pi$.
All models were run with 100 particles each for 5 different $\beta$s: 
0.01, 0.05, 0.1, 0.2 and 0.4; for a density of 2.7 g cm$^{-3}$, 
these values correspond to particle sizes of 40, 9, 4, 2 and 1 $\mu$m, 
respectively; for  1 g cm$^{-3}$, they correspond to 120, 23, 11, 
6 and 3 $\mu$m, respectively.

Figure 3 shows the lifetimes for all the particles in our models.
For the present discussion, the difference between having 4 or 7 planets is
not important. What is important for this argument
is that the initial conditions of the parent bodies are different upon 
release of the dust particles. 
We see in Figure 3 that the median lifetime and the dispersion of lifetimes are
both systematically smaller for larger $\beta$. The longer lifetimes are
due to longer trapping at exterior MMRs with Neptune. The residence time in an 
MMR is variable and unpredictable owing to the underlying chaotic resonance dynamics
(cf. Malhotra et al.,~\citeyear{malh00}). From the point of view of using
these simulations to obtain the equilibrium spatial distribution of dust, the
most worrisome feature is that the lifetime of the longest-lived
particle may be several times longer than the next longest-lived, and more
than an order of magnitude greater than the median lifetime. This may be due to 
numerical errors that affect the behavior of a few particles, or it 
may be due to the underlying chaotic dynamics that produces a long
tail in a continuous distribution of dynamical lifetimes. 

To distinguish between these two possibilities, two additional runs of
100 particles each (with different random values of M, $\Omega$ and $\omega$)
were done for Model I-A with $\beta$=0.1. The results are shown in Figure 4.
We see that with increasing number of particles, the gap between the longest
and next longest lived particle is reduced. Overall, the distribution of lifetimes
resembles the sum of a gaussian and a uniform distribution. With only
a few hundred particles in numerical simulations, we are limited to 
small numbers of long lived particles. However, we conclude with some
confidence that the longest lived particles are not anomalous but statistically
representative of a real dynamical population.

\subsection{Spatial Distribution}
Figure 5 shows the ``equilibrium'' number density distributions that 
result after applying LZ99 approach to the three models I-A with $\beta$=0.1. 
The relative occurrence of the different MMRs can also be seen in the 
histogram presented in this figure. Note that the only difference between the three
runs is in the initial M, $\Omega$ and $\omega$. 
We see that the dust particles' times of residence in various mean resonances 
with Neptune are highly variable.

Figure 6 shows the radial profiles 
(averaged over all $\theta$) and angular profiles (integrated 
between 25 and 35 AU) of the number and brightness (see $\S$5) 
density distribution derived from 4 different sets of 100 particles each.
This figure indicates
that the LZ99 approach: (1) is able to predict reliably the radial structure;
and (2) the  azimuthal structure is not predictable in detail, 
except for a `gap' near the outermost planet Neptune. 

We have explored how fast structure is created and the effect of 
excluding the contribution to the structure from the longest lived 
particles. Our results, which are summarized in Figure 7, show that the structure is 
created quickly and that the radial profiles of the number density
distribution do not strongly depend on the contribution from the 
longest lived particles. This provides further validation of the LZ99 approach.

\section{Distribution of Parent Bodies and Orbital Evolution of Dust Particles}
\label{dist}
KBOs are icy bodies that lie in a disk beyond Neptune's orbit. It 
is estimated that there are about 3.5$\cdot$10$^4$ objects with
diameters $>$ 100 km (Jewitt \& Luu,~\citeyear{jewi95}) in the 30-50 AU
annulus. The outer limit of the belt is presently not well determined.
Dust production occurs due to mutual collisions of KBOs 
(Backman \& Paresce,~\citeyear{back93}; 
Backman, Dasgupta \& Stencel,~\citeyear{back95}; Stern,~\citeyear{ster96}) and 
to collisions with interstellar grains (Yamamoto \& Mukai,~\citeyear{yama98}).

Our selection of the orbital elements of the parent bodies is based on 
published observations of KBOs and on 
recent studies of their debiased radial
(Trujillo \& Brown,~\citeyear{truj01a}) and inclination distributions
(Brown,~\citeyear{brow01}). 
Semimajor axis were uniformly distributed between 35 and 50 AU; 
eccentricities were derived from perihelion distances, with random values
between 35 and 50 AU; inclinations were uniformly distributed between 0 and 17$^\circ$,
and the other three orbital elements, mean anomaly 
(M), longitude of ascending node ($\Omega$) and  argument of perihelion 
($\omega$), were randomly selected between 0 and 2$\pi$.

When dust particles are released from their parent bodies ($\beta$=0), 
their orbital elements instantaneously change due to the effect of 
radiation pressure that, as we saw in $\S$2.2, makes the particle 
$\it{feel}$ a less massive Sun by a factor (1-$\beta$). 
Their new semimajor axis ($\it{a'}$) and eccentricity ($\it{e'}$) in terms
of their parent bodies' ($\it{a}$ and $\it{e}$) are given by
\begin{equation}
{a'=a{1-\beta \over 1-2a\beta/r}}
\end{equation}
\begin{equation}
{e'={\mid 1 - { (1-2a\beta /r)(1-e^2) \over (1-\beta^2)}\mid}^{1/2}.}
\end{equation}
Figure 8 shows $\it{e}$ and $\it{i}$ for the parent bodies and the
dust particles at the time of release.

In their slow journey towards the Sun, the particles cross
MMRs with the giant planets. As a result, some particles get trapped 
and structure in the debris disk begins to form.
As reported by LZ99, 
and also seen in our models, the exterior resonances with Neptune 
dominate the trapping. Usually, the particles escape the resonances via 
close encounters with the planet, but in the case of interior resonances, 
they can also escape due to the decrease of $\it{a}$, that makes the particle
get farther away from the planet where drag forces dominate (Liou \& Zook,~\citeyear{liou97}). 

We have used the three models I-A with $\beta$=0.1 to study the existence 
of correlations in the initial orbital elements of the longest-lived 
particles. Figure 9 shows $\it{a}$, $\it{e}$, $\lambda$-$\lambda_{Neptune}$ and M
for the 65 longest lived particles (solid lines; these particles have 
lifetimes $\ge$2$\cdot$10$^{7}$ years; see Figure 4) compared to 
all the 300 particles in the models (dotted lines).  
There are two prominent features both readily understood:
(1) As the particles are released, and due 
to their increased semimajor axis, their mean anomaly is such that
they avoid aphelion, explaining the gap between 90$^{\circ}$ and 
270$^{\circ}$. (2) The longest-lived particles tend to have smaller initial
eccentricities, as expected from the fact that they tend to be trapped
more easily in resonances. We find no evidence of correlation between
lifetime and initial orbital parameters.
 
\section{Structure Formation: the Giant Planets Reshape the Debris Disk}
\label{struc}
Figure 10 shows the equilibrium semimajor axis distributions,  
Figures 11 and 12 show the equilibrium number
density distributions in the presence and absence of planets, and Figure 13 shows the radial 
profiles averaged over all $\theta$. The main features seen in these figures are: 
(1) the ring-like structure along Neptune's orbit, showing some azimuthal variation due to MMRs; 
(2) the minimum density at Neptune's position, as particles 
in MMRs tend to avoid the perturbing planet; (3) the clearing of dust from the inner 
10 AU; and (4) the fact that the structure is more prominent for larger 
particles (smaller $\beta$s).  The latter is because the trapping in MMRs is 
more efficient when the drag forces are small (LZ99). On the other hand, the ejection of
particles from the inner 10 AU does not depend on size.  
The difference between models I-A and II-A in Figure 10 gives
an estimate of the uncertainties, since the effect of the 3 terrestrial 
planets is negligible and the only difference is in the initial conditions 
of the parent bodies. The relative ``strength'' of the dominant MMRs 
depends quite strongly on the initial conditions (see also the histogram in 
Figure 5).
This may indicate that the exact prediction of a planet's orbit, 
based on the identification of resonances,
may be difficult.
The ring-like structure in the number density is 
also visible in the brightness distributions of Figure 11, which were calculated 
assuming greybody absorption and emission by the dust grains in a 3$\cdot$10$^{-11}$\msol~ 
single size grain disk, at a distance of 30 pc. 
Additional features seen in the brightness distribution are: (1) a bright ring 
between 10 and 15 AU with a sharp inner edge, due to the ejection of 
particles by Saturn and Jupiter; and (2) a steep increase in brightness 
in the inner 5 AU. Both features are the combination of the decreasing particle density and 
increasing grain temperature closer to the Sun.

>From the observational point of view, current 
IR detector technology does not allow us to spatially resolve many of these 
features. As an example, the SIRTF MIPS 24 $\mu$m detector has a pixel 
size of 2.45'', that at the distance of $\beta$-Pictoris (16.4 pc) 
means a spatial resolution of 40 AU. For SITRF IRAC (3.6-8.0 $\mu$m) 
the resolution would be about 20 AU. The question is then how much 
information can be derived from the disk SED.
Figure 14 shows the compositive SEDs that result from combining the 
SEDs from the $\beta$=0.01, 0.05, 0.1, 0.2 and 0.4 disks, with weights in such a way that 
they follow the power law distribution 
n($\it{a}$)d$\it{a}$=n$_0{\it{a}}^{-3.5}$d$\it{a}$, where 
$\it{a}$ is the particle radius.
The black lines correspond to the SEDs from a 3$\cdot$10$^{-11}$\msol~disk.
Blue, red and green correspond to the SEDs from a system of Sun plus a 
disk with three different masses. In all cases, the solid line is for a system 
with 7 planets and the dotted line is for a system without planets. 
The wavelength labels correspond to the SIRTF MIPS and IRAC bands potentially
useful to study these systems. 
We see that the presence of planets does modify the disk SED.
The main modification is due to the clearing of dust in the inner region
(an ``inner gap'') by Jupiter and Saturn, which causes a significant deficit 
in the disk SED at higher frequencies. The density enhancement in the annulus 
between 35 and 50 AU, due to trapping in Neptune's exterior MMRs,
causes a relatively smaller effect on the shape of the disk SED. 
How well can one determine the masses and orbits of planetary perturbers 
from the shape of the disk SED?  We plan to address this question in the 
future by exploring in detail the parameter space of planetary masses and 
orbital elements.

It is important to note that our model systems (with and without planets)
contain the same amount of disk mass.  We are interested in how the structure 
created by the planets affects the shape of the SED, independent of the dust 
production rate.  The latter determines only the normalization factor. 
However, planetary perturbations can affect the dust production rate, 
possibly leading to more massive dust disks.  This effect is not taken into 
account in our models, but will be considered in the future.

\section{Dust destroying processes}
\label{destroy}
\subsection{Collisions}
Particles that from the dynamical point of view are able to drift all the way
into the Sun, may get destroyed by mutual collisions or collisions 
with interstellar dust grains before they reach the inner Solar System.
Based on Ulysses measurements of interstellar dust flux at 5 AU, 
and assuming that this flux is constant throughout the Solar System and 
does not vary in time, the average time for
one collision to occur between an spherical grain of diameter $\it{d}$  and  an 
interstellar grain of diameter $\it{d}_i$: $\it{t}_c$=504/($\it{d}$+$\it{d}_i$)$^2$
Myrs (LZD96).  Assuming that interstellar dust have an average size of 1.2 $\mu$m,
the collisional times for 1,2,4 and 9 $\mu$m~ particles are 
104, 49, 19 and 4.8 Myrs respectively. For densities of 2.7 gcm$^{-3}$ 
these sizes correspond to $\beta$s of 0.4, 0.2, 0.1 and 0.05 respectively.
KB dust, however, are more likely to have lower densities. Analysis of 
collected IDPs indicate that high velocity IDPs have fluffy, porous textures 
with  an average density of 
about 1 gcm$^{-3}$ (Joswiak et al.,~\citeyear{josw00}). For those densities
the sizes corresponding to the $\beta$s above are 3, 6, 11 and 23 $\mu$m.
These particles will have collisional times of 28.6, 9.7, 3.4 and 0.86 Myr
respectively.
In these size ranges mutual collisions are not as important 
as collisions with interstellar grains (LZD96). If so, comparing 
the collisional times and the dynamical lifetimes in Figure 3 
shows that collisional destruction is only 
important for grains larger than about 6 $\mu$m. Smaller particles will 
therefore survive collisions and drift all the way into the Sun contributing to the 
zodiacal cloud. Particles larger than 
50$\mu$m may also survive collisions because interstellar 
grains are too small to destroy these in a single impact, so 
it is possible that they are able to evolve into the inner Solar System
(LZD96). 
Figure 7 shows the timescale for disk structure formation in the case of 
$\beta$=0.1. Structure is already beginning to form by about 8 Myrs; 
by 16 Myr, the structure shows almost all the features of the equilibrium
state. Collisional time scales for $\beta$=0.1 range from 3.4 to 19 Myrs, 
depending on the density. It is not clear therefore that disk structure for these
particles sizes is able to survive collisions. For smaller particles (larger
$\beta$s) structure will survive, but these particles do not have as prominent a
structure associated with the exterior MMRs with outer planets (see Figure 5).
Although all these results should be taken with caution, since 
the flux and the size distribution of the interstellar grains are rather 
uncertain, what is clear is that one should keep in mind collisions 
with interstellar grains when trying to infer the presence of planets from 
the study of structure in debris disk (see also LZ99).

\subsection{Sublimation}
Depending on the composition of dust particles, sublimation may or may not 
play an important role in dust destruction processes and therefore in the
ability of dust to reach the inner Solar System. For silicates, the sublimation 
temperature is $\sim$1500 K. For the particles sizes considered in this
paper, 1, 2, 4, 9 and 40 $\mu$m (that correspond to
$\beta$s of 0.4, 0.2, 0.1, 0.05 and 0.01 with $\rho$=2.7 gcm$^{-3}$), 
this temperature is reached at 
r$<$0.5 AU, which is the minimum heliocentric distance allowed by our models.  
In this case, sublimation does not affect the evolution of dust particles
and the radial disk structure. But if the KB dust composition is more 
similar to water ice, the sublimation temperature is $\sim$100 K, that 
for the sizes of 3, 6, 11, 23 and 120 $\mu$m (corresponding to the $\beta$s 
above with $\rho$=1 gcm$^{-3}$), is reached at 27, 19, 14, 10 and 4.3 AU 
respectively.  In this case, the ability of dust to reach
the inner Solar System would be greatly diminished by sublimation, 
even for dust grains as large as 120 $\mu$m, and the disk structure created by
the inner planets would be destroyed.

\section{Conclusions and Future Work}
\label{concl}
(1) We have followed, from source to sink, the orbital evolution of dust particles 
from the Kuiper Belt. To integrate the equations of motion efficiently, 
we have introduced radiation and solar wind forces in 
the multiple time step symplectic integrator of DLL98.
We have established the suitability of our code by 
comparison between numerical results and analytical solutions to 2-body and
restricted three-body cases, as well as comparison with other numerical
results in the literature (LZD96, LZ99). 

(2) We have carried out numerical simulations for single size particle 
disks in the presence and in the absence of planets in order to 
estimate the uncertainties inherent in the 
prediction of structure in the outer solar system debris disk, owing 
to the chaotic dynamics of dust orbital evolution.  We simulate dust particle
initial conditions according to the wider distribution of parent bodies
indicated by the recent observed distribution of KBOs, and our simulations 
extend to larger particle sizes than previous studies. 

(3) We find that the distribution of KB dust particle lifetimes in the
Solar system are described as a sum of a gaussian and a nearly uniform
distribution; the latter represents only a small fraction of all particles
but extends to very long lifetimes, while the gaussian represents the 
dominant fraction of particles. The mean and dispersion of the gaussian
component increases systematically with particle size, and is in the range
of [few million years] for [1--100 $\mu$m] particle sizes.
We do not find any correlations between the initial orbital elements 
and dynamical lifetimes of dust particles. 

(4) We have examined carefully the method used by LZ99 to estimate the
equilibrium spatial distribution of KB dust in the Solar System.
This method is based on the ergodic assumption, so the dust structures 
obtained are determined to a large extent by the longest lived particles,
which represent only a very small fraction of the dust population. 
The ergodic assumption is generally not applicable in chaotic dynamical systems.
Nevertheless, we have established that in practice this method gives reliable
results for several aspects of dust dynamical studies for three reasons: 
(i) the distribution of dust particle lifetimes is described as a sum 
of a gaussian plus a nearly uniform distribution, i.e. the longest-lived 
particles are not anomalous, they are statistically representative of 
the long tail population; 
(ii) the dust spatial structure is created quickly; 
(iii) the radial profile of the equilibrium number density distribution does not 
strongly depend on the longest-lived particles (although the azimuthal structure 
does). 

(5) Overall, the number density of the KB dust disk shows a depletion of 
dust in the inner 10 AU, due to gravitational scattering by Jupiter and Saturn, 
and an enhanced dust density in a ring between 35 and 50 AU, due to trapping of 
particles in MMRs with Neptune. The structure is more pronounced for larger 
particle sizes. The brightness distribution shows a bright ring between 
10 and 15 AU with a sharp inner edge (particles ejected by Saturn and Jupiter), 
and a steep increase in brightness in the inner few AU (a combination of the 
decreasing density and increasing grain temperature). 

(6) We find that the azimuthal structure of the dust disk is not predictable in 
detail, except for a `gap' near the outermost planet Neptune.  This is because 
the azimuthal structure depends sensitively on the long lived particles trapped 
in mean motion resonances with Neptune, and the times of residence in the 
various resonances are highly variable and unpredictable.

(7) We have calculated disk brightness density and spectral energy
distributions (SED), assuming greybody absorption and emission from the dust
grains. 
We find that the presence of planets modifies the shape of the SED.
The Solar System debris disk SED is particularly affected by the clearing of 
dust from the inner 10 AU due to gravitational scattering by Jupiter and 
Saturn. 

(8) Grain physical lifetimes are limited by collisions and sublimation.
The comparison of the dynamical lifetime of particles, the timescale
for structure formation and the collisional time between KB and interstellar
grains indicates that, if the current estimates for the flux and the size
distribution of interstellar grains are correct, collisional destruction is
important for grains larger than about 6 $\mu$m. For smaller particles,
debris disk structure will be able to survive, although the smaller particles
have less prominent structure associated with the outer planets.
Depending on their composition, sublimation of 
particles may or may not play an important role in the destruction of 
structure. If KB dust has water ice composition, and assuming a sublimating
temperature of 100 K, it is likely that even large 120 $\mu$m particles will 
sublimate before reaching the inner 4 AU of the Solar System. 
We conclude that grain destruction processes need to be examined more carefully
in future applications of our studies to infer the presence of planets 
from structure in debris disks.

This work is part of the SIRTF FEPS Legacy 
project\footnote{http://feps.as.arizona.edu} (P.I. M. Meyer), with 
the goal ``to establish the diversity of planetary architectures 
from SEDs capable of diagnosing the radial distribution of dust and the 
dynamical imprints of embedded giant planets''.  
The modeling of a particular system is very complex, because it involves a 
large number of free parameters. We have therefore chosen a forward modeling 
approach: a grid of models will be created for different planetary masses and 
orbital radii, parent bodies' masses and orbital distribution, total mass in dust 
particles, etc.  
We will produce dust spatial distributions like the ones presented here which 
will be used as input for a radiative transfer calculation to generate SEDs 
containing all the important spectroscopic features.  This will be more detailed 
than the simple greybody approximation used in the present work.
This ``library'', that as part of our Legacy will be available to 
the community, will contain the templates to which we will compare the dust 
SEDs derived from the SIRTF observations for their 
interpretation in terms of planetary architectures.

\begin{center} {\it Acknowledgments} \end{center}
We thank Hal Levison for providing the SKEEL computer code.  
AMM is supported by NASA contract 1224768 administered by JPL. 
RM is supported by NASA grants NAG5-10343 and NAG5-11661.

\clearpage

\begin{figure}\label{f1}
\epsscale{0.95}
\caption{
Evolution of $\it{a}$ and $\it{e}$ for a particle
with $\beta$=0.2 and sw=0.35 in the 2-body problem. 
The solid and the dotted lines coincide and represent the numerical
and analytical results respectively.}
\end{figure}

\begin{figure}
\epsscale{0.67}
\caption{
Comparison between numerical (solid line) and analytical results
for the the evolution of the orbital elements of
1\mum~ pyroxene dust particles ($\beta$=0.17, sw=0.35) in a circular
Sun-Neptune-dust system. Neptune is placed at 30 AU with $\it{e}$=0.
The dotted lines represent the analytical results given by
(19) and (20) (valid to 2nd order in eccentricity and inclination)
and the dashed line correspond to (18) (valid for all
eccentricities when i=0).
(a) Particle trapped for 2 Myr in the 1:1 MMR with Neptune. Since
the eccentricity is already quite large at the time of trapping (e$\sim$~0.3),
the agreement with (19) is not very good. (b) Particle trapped for
50 Myr in the exterior 5:3 MMR with Neptune. The agreement is very good until
the eccentricity reaches $\sim$~0.3, at that point it starts to deviate.
(c) Particle trapped for 14 Myr
in the exterior 4:3 MMR with Neptune. At the time of trapping, the
inclination is very small ($\sim$~0.6$^\circ$). The evolution of the
eccentricity is perfectly described by (18) and (19).
(d) Particle trapped for 4 Myr in the interior 5:6 MMR with Neptune.
The overall evolution of eccentricity and inclination are described
reasonably well by (19).
The semimajor axis stays constant as the eccentricity
decreases until it reaches the limiting value 0, the point at which the
particle leaves the resonance.}
\end{figure}


\begin{figure}
\epsscale{0.8}
\caption{Lifetimes of the particles in models I-A (with 4 planets; black solid line),
models I-B (without planets; black dotted lines), models II-A
(with 7 planets; red solid line) and models II-B (without planets;
red dotted lines). The insert for
$\beta$=0.01 is included to show the full time expand of these very
long-lived particles. The inserts for $\beta$=0.05 and $\beta$=0.1
show the no-planet models separately to avoid confusion.
The presence of the planets increases the lifetime of the particles.
The smaller the beta, the largest the difference between the planets
and no-planets cases: the trapping into MMRs is more efficient when the drag
force is small.}
\end{figure}


\begin{figure}
\caption{Lifetimes for the three models I-A, 100 particles each, plotted together with
the different colors representing the contribution from the 3 different
runs. The distribution of lifetimes reassembles that of a gaussian (dotted
line) plus a long tail.}
\end{figure}


\begin{figure}
\caption{
(a), (b) and (c) ``Equilibrium'' number density distributions for the
three  models I-A with $\beta$=0.1. (d) Number density
distributions for the 300 particles together. (e) 10$^{5}$ randomly
selected points from (a) indicating that a large number of particles
is needed to resolve the structure. The dot indicates the position of
Neptune. The histogram shows the relative
occurrence of the different MMRs. The position of a few MMRs with Neptune
are indicated in the figure.}
\end{figure}


\begin{figure}
\caption{
(top) Number and brightness density radial distributions, averaged over
all $\theta$, for particles with $\beta$=0.1. Black, red and blue correspond
to the three models I-A. Green corresponds to the
model II-A.
(bottom) Same as above but for angular distributions, integrated
between 25 and 35 AU. The longitude is measured with respect to
Neptune.  LZ99 approach is able to predict the
radial structure, but the uncertainties in the azimuthal structure are
large.}
\end{figure}


\begin{figure}
\caption{
(top) Time scale in which structure is created: (a) Number density
distribution for one of the models I-A with $\beta$=0.1
by the time the last particle leaves the system (125.9 Myr). (b), (c) and
(d) show the structure seen at earlier and earlier times: 31.8, 15.9
and 7.95 Myr, respectively. (bottom) Effect of excluding the longest-lived
particles: (e), (f), (g), (h) show the structure
after excluding the contribution from the 2, 6, 10 and 14 longest-lived
particles respectively.  These results validate the use of LZ99 approach
by indicating that the structure is
created quickly and that the radial profiles of the number density
distribution do not strongly depend on the contribution from the
longest-lived particles. The dot indicates the position of Neptune.}
\end{figure}


\begin{figure}
\caption{
Distribution of eccentricities and inclinations for parent bodies
(green), dust particles at the time of release (red), evolved
dust particles in models II-A (blue) and evolved
dust particles in models II-B (black).
The difference between the
presence and non presence of planets is more dramatic for smaller $\beta$s.
When planets are present, a fraction of the particles have their
eccentricities and inclinations increased (due to trapping in
$\it{e-type}$ and $\it{i-type}$ exterior resonances respectively).
Radiation forces do not affect inclination, so the green, red and black
lines coincide.}
\end{figure}


\begin{figure}
\caption{
Initial orbital elements of the 65 longest-lived particles
from the three  models I-A with $\beta$=0.1 (solid line),
compared with the total of 300 particles (dotted line).
The longest-lived particles tend to have lower
$\it{e}$. The gap between 90$^{\circ}$ and 270$^{\circ}$ is explained
because upon release, due to the increased $\it{a}$, the particles
avoid aphelion.}
\end{figure}


\begin{figure}
\caption{
``Equilibrium'' semimajor axis distribution in logarithmic scale for
the particles in the models I-A (black solid lines), models I-B
(black dotted lines), models II-A
(red solid lines) and models II-B (red dotted lines).
The trapping of particles in the exterior MMRs with Neptune and the
depletion of particles in the inner 10 AU in the presence of planets
are the most prominent features in the figure.}
\end{figure}


\begin{figure}
\caption{
``Equilibrium'' number density distribution for models I-A and II-A
(columns 1 and 2) and brightness density distribution for model II-A
(column 3). The brightness density is in units of
ergs$^{-1}$cm$^{-2}$(1AU)$^{-2}$ and
corresponds to the thermal emission, integrated from 21.6 to 26.3 $\mu$m,
of a 3$\cdot$10$^{-11}$\msol~ disk at a distance of 30 pc surrounding a
1 \lsol~ star. Grain temperatures were calculated using the expressions in
Backman \& Paresce (\citeyear{back93}) for the thermal equilibria and
emitted spectra of generic grains. Absorptive efficiency was assumed to
be $\epsilon$=1 and emissive efficiency was $\epsilon$=1 for
$\lambda<\it{a}$ and $\epsilon$=$\it{a}/\lambda$ for $\lambda>\it{a}$,
where $\it{a}$ is the grain radius. The dust particles have $\rho$=2.7
gcm$^{-3}$. The dot at (30,0) indicates the position of Neptune.}
\end{figure}


\begin{figure}
\caption{
Same as Figure 11 but for the ``equilibrium'' number and brightness
density distributions of models II-B.}
\end{figure}


\begin{figure}
\caption{
Number and brightness surface density radial distributions, averaged over
all $\theta$, for models II-A (top) and
models II-B (bottom) shown in Figures 11 and 12.
The main features are the depletion of particles in the inner 10 AU,
due to scattering by Jupiter and Neptune, and the enhancement
of particles from 30 to 50 AU, due to trapping in MMRs with Neptune.}
\end{figure}


\begin{figure}
\caption{
(top) Compositive SEDs that result from combining the SEDs from the 
$\beta$=0.01, 0.05, 0.1, 0.2 and 0.4 disks, with weights in such a way 
that they follow the size distribution 
n($\it{a}$)d$\it{a}$=n$_0{\it{a}}^{-3.5}$d$\it{a}$, where $\it{a}$
is the particle radius. $\it{Black}$ is for a 3$\cdot$10$^{-11}$\msol~disk 
only;
$\it{blue}$ is for Sun + 3$\cdot$10$^{-11}$\msol~ disk; $\it{red}$ is for Sun +
3$\cdot$10$^{-10}$\msol~ disk; and $\it{green}$ is for Sun +
3$\cdot$10$^{-9}$\msol~disk. In all cases, the solid line is for a system with
7 planets, the dotted line is for no planets and
the system is at a distance of 30 pc. 
(bottom) Same as top but in Jy vs. $\mu$m.
The squares correspond to the data points, indicating the spectral resolution
on the synthetic SEDs.}
\end{figure}

\end{document}